*FPCB : a simple and swift strategy for mirror repeat identification*

Bhardwaj Vikash*[1], Gupta Swapnil[2], Meena Sitaram and Sharma Kulbhushan*[2]

[1]*Government College Sector 14 Gurgaon, Haryana, India.*

[2]*Division of Radiation Biosciences, Institute of Nuclear Medicine and Allied Sciences, Delhi -110054, India.*

**Running Title:** Three step strategy for mirror repeat identification.

*Corresponding authors:

1. Kulbhushan Sharma, Metabolic Cell Signaling Group

Division of Radiation Biosciences,

Institute of Nuclear Medicine and Allied Sciences, Delhi -110054, India

Phone: 91-11-23905186; Fax: 91-11-23919509

E-mail: kulsinmas@gmail.com

2. Vikash Bhardwaj, Government College Sector 14 Gurgaon, Haryana, India.

Email: vikashbhardwaj@ gmail.com.

## ABSTRACT

After the recent advancement of sequencing strategies, mirror repeats have been found to be present in the gene sequence of many organisms and species. This presence of mirror repeats in most of the sequences indicates towards some important functional role of these repeats. However, a simple and quick strategy to search these repeats in a given sequence is not available. We in this manuscript have proposed a simple and swift strategy named as FPCB strategy to identify mirror repeats in a give sequence. The strategy includes three simple steps of downloading sequencing in FASTA format (F), making its parallel complement (PC) and finally performing a homology search with the original sequence (B). At least twenty genes were analyzed using the proposed study. A number and types of mirror repeats were observed. We have also tried to give nomenclature to these repeats. We hope that the proposed FPCB strategy will be quite helpful for the identification of mirror repeats in DNA or mRNA sequence. Also the strategy may help in unravelling the functional role of mirror repeats in various processes including evolution.

## INTRODUCTION

After completion of DNA sequencing, many different types of sequences able to have a regulatory role have been discovered. Among these sequences mirror repeats have been predicted to play an important roles like adoption of perfect or near-perfect homopurine or homopyrimidine mirror repeats into triple-helical H conformations (Spano *et al*, 2007) . A mirror repeat is for example 5'AGTTCATTACTTGA3'where the sequences AGTTCAT and TACTTGA are mirror repeats of each other (Fig 1). The repeats may or may not be separated by a spacer nucleotide sequences in between them. There are several computer programs available to develop to detect repeats and/or the associated secondary structure in DNA or RNA sequences. However, most of these programs have not been able to effectively find out mirror repeats.

The purpose of this work is to devise a simple strategy to find out a mirror repeat in a sequence. We have recently shown through carious techniques that parallel DNA synthesis is quite possible in PCR reaction when we introduced a parallel primer in the reaction (Bhardwaj *et al,* 2013). The interesting outcome in this reaction is that the final product resulting from parallel DNA PCR (PD-PCR) is the original template DNA in reverse orientation so that the initial 5' end becomes 3' end. We have shown in this work that if we align parallel complement of a nucleotide sequence to original nucleotide sequence and perform a blast analysis, mirror repeats present in original nucleotide sequence can be detected easily. We have analyzed twenty genes to validate our strategy. During this, we found the presence of various types of repeats in nature. We have tried to give nomenclature to these mirror repeats.

One of the problems in counting repeats is the fact that a single repeat can be counted many times if one does not define in some way a maximal repeat (Spano *et al*, 2007). When one considers mirror repeats, the definition of maximal repeat is less clear and must be clearly defined. We have also tried to define the same in this work. We are interested in finding the numbers and types of mirror repeats in model genome sequences. Interesting, no novel software is required for the proposed strategy. We just have to follow three simple steps for the identification of mirror repeats in a given sequence.

We hope that this simple strategy will be helpful in identifying mirror repeats in a sequence. Also, the strategy will be helpful in understanding the role of these mirror repeats in a sequence and will also assist in understanding the events of evolution in terms of mirror repeats.

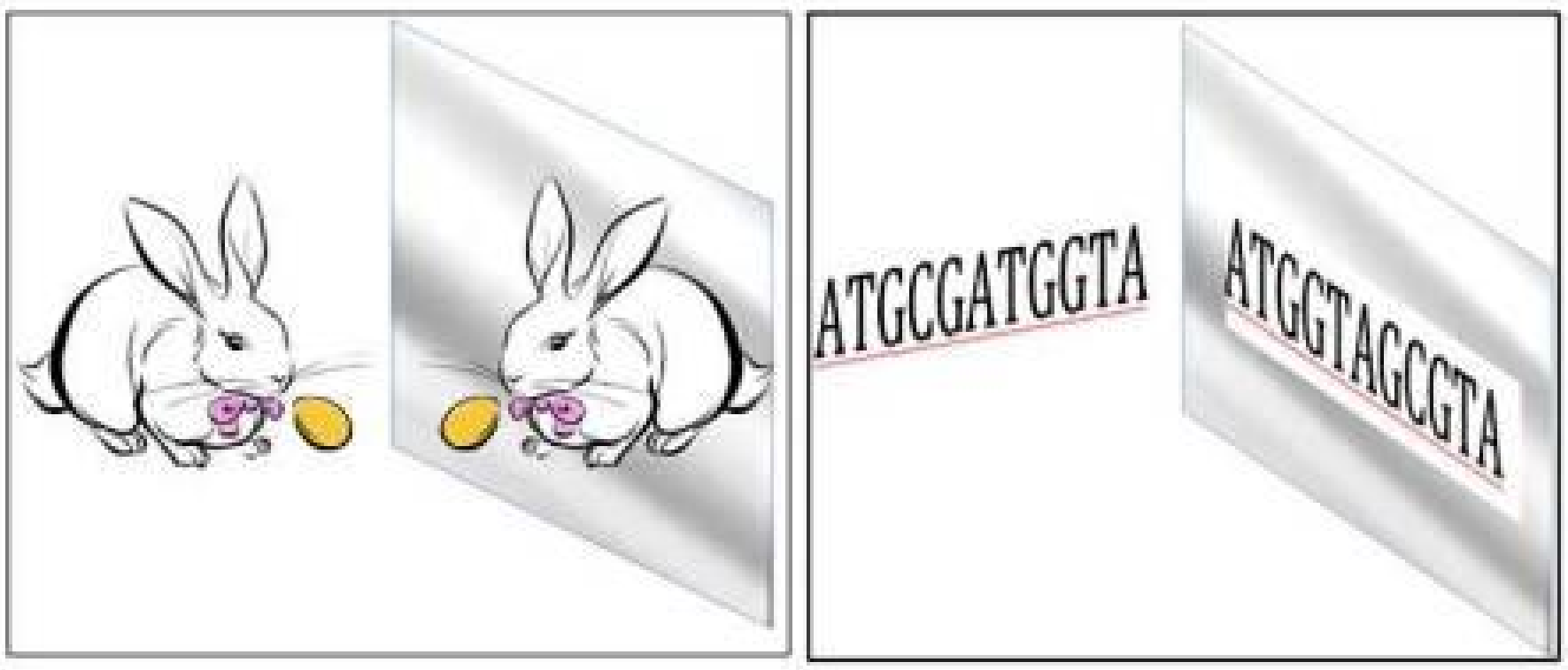

**Figure 1: Mirror repeats :** DNA mirror repeat (MR) is a sequence segment delimited on the basis of its containing a centre of symmetry on a single strand and identical terminal nucleotides. Picture in left depicts the mirror image of an animal whereaspicture on right depicts an example of a nucleotide sequence mirror repeat.

## MATERIALS and METHODS

1. *Downloading sequence in FASTA format*

The coding sequences for the gene of interest were downloaded in FASTA format using the following link: http://www.ncbi.nlm.nih.gov.

2. *Making a Parallel complement*

Once downloaded, the FASTA format of nucleotide sequences were pasted into Reverse Complement program (http://www.bioinformatics.org/sms/rev_comp.html) and DNA sequences were converted into its complement counterpart.

3. *Mirror repeat search*

Both FASTA format of Nucleotide sequence and its parallel complement were aligned for BLAST homology search using BLAST tool: (http://blast.ncbi.nlm.nih.gov/Blast.cgi?PAGE=MegaBlast&PROGRAM=blastn&BLAST_PROGRAMS=megaBlast&PAGE_TYPE=BlastSearch&BLAST_SPEC=blast2seq&QUERY=&S

UBJECTS=). The program was optimized for somewhat similar sequences (blastn) which allows a word size down to seven bases. If the position number is exactly reversed in subject and query , it will be a mirror repeat.

4. *Gene analysis and Nomenclature of mirror repeats*

Twenty important mammalian genes were analysed using the proposed study. The nomenclature was proposed for different types of mirror repeats observed during the analysis.

**RESULTS**

A simple strategy named **FPCB (FASTA-parallel complement-BLAST)** was devised to find out mirror repeats in various human mRNAs. The coding sequences for the gene of interest were downloaded in FASTA format using http://www.ncbi.nlm.nih.gov. In order to obtain the reverse complement sequence, the sequence in FASTA format was pasted into the following program (http://www.bioinformatics.org/sms/rev_comp.html). The final output of reverse complement program was selected for homology search with the parental sequence. Complement sequence was aligned with original sequence was submitted for BLAST analysis. The FPCB strategy has been depicted in Figure 2.

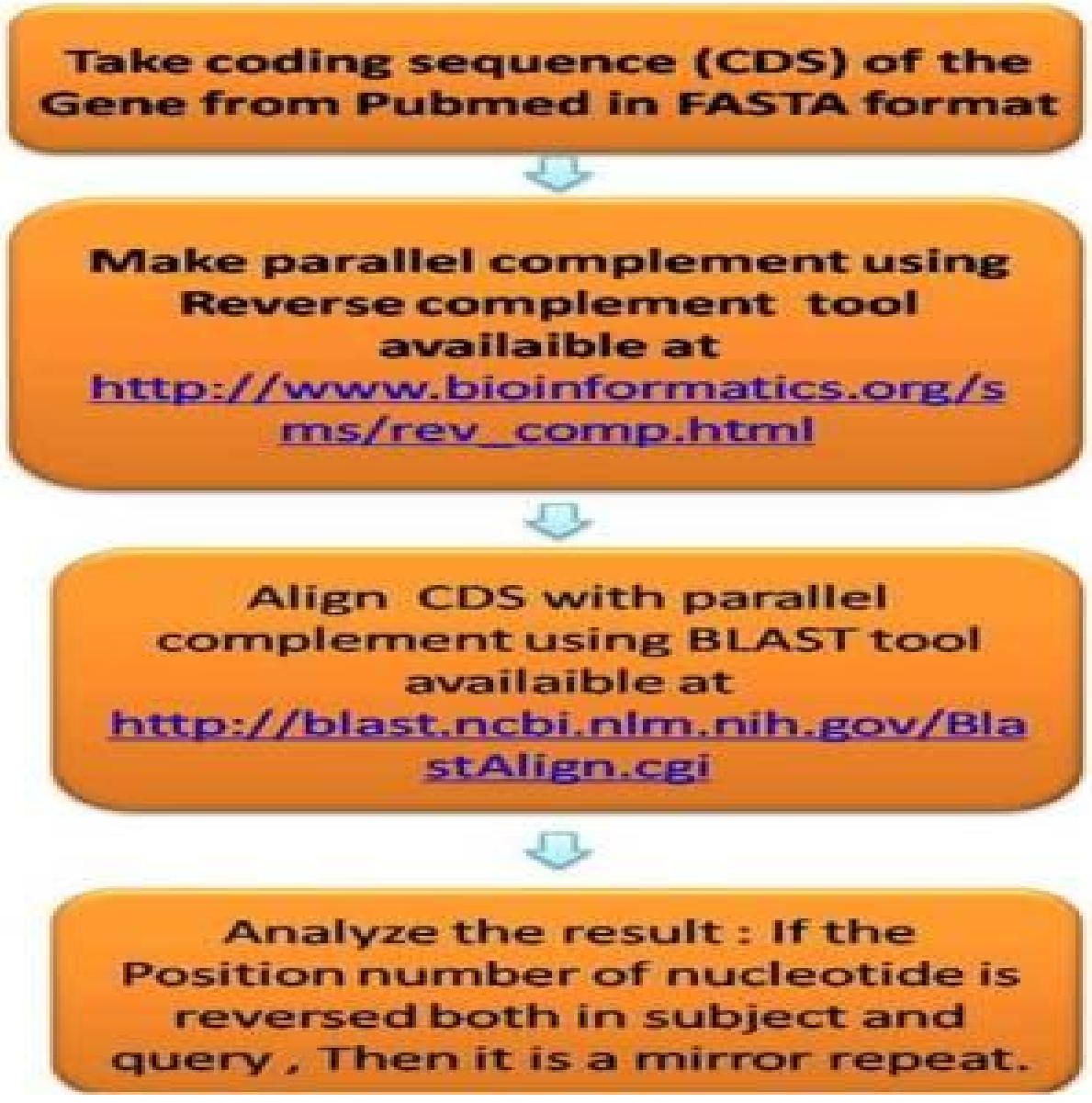

**Figure 2: FPCB Strategy to find out mirror repeats.** The strategy is based on three simple steps reflected in the name assigned FPCB. F stands for the first step of downloading gene coding sequence in FASTA format. PC stands for formation of parallel complement of the gene of interest. The third and final step of the strategy is BLAST alignment analysis depicted by the word B.

The strategy was applied for twenty genes. Various genes selected for the FPCB analysis has been shown in table 1. During this analysis, a number and types of mirror repeats were observed. We have proposed the nomenclature of these mirror repeats (Fig.3). A typical mirror repeat has been shown in Fig.3 (i). Depending on the number of spacers at centre of symmetry of the mirror repeats, these were named as single spacer mirror repeats (SSMR), double spacer mirror repeats (DSMR) and multi spacer mirror repeats (MSMR) (Fig.3, ii, iii and iv respectively). Tandem mirror repeats and continuous di-mirror repeats were also observed (Fig.3, v and vi respectively). In some of the genes, few rare and interesting mirror repeats were observed. These were Continuous overlapping mirror repeats (COMR) and mirror repeats with simple tandem repeats (MRSMR).

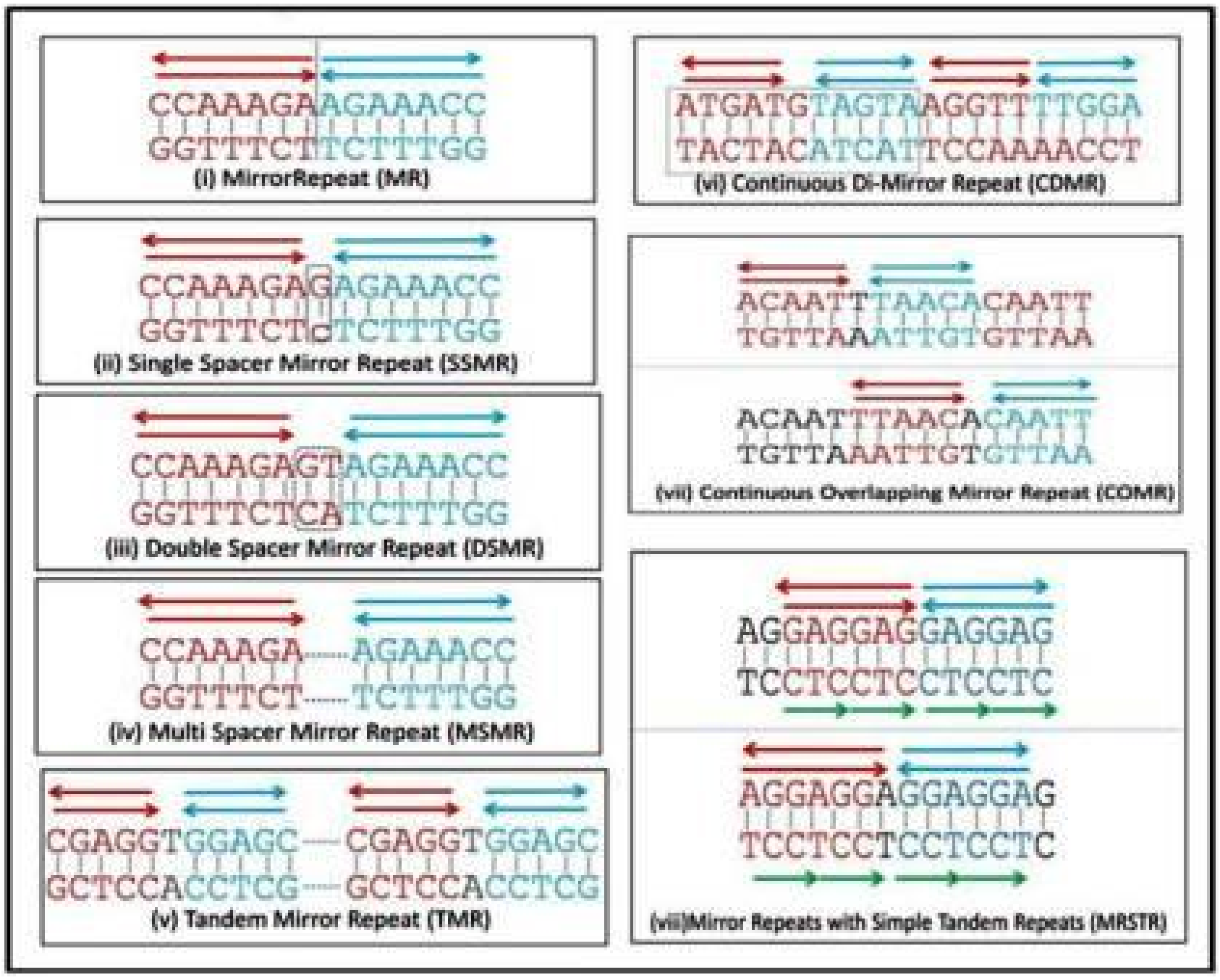

**Figure 3: Nomenclature and analysis of various types of mirror repeats observed after the use of FPCB Strategy.** At least eight different types of mirror repeats were obtained after FPCB analysis in different genes. The strategy is based on three simple steps reflected in the name assigned FPCB.

Various important human genes were arbitrarily selected. Some of these genes included p53, BRCA 1, COX-2, NFkB, TGF beta etc. (Table 1). We included only mammalian genes in this study. However, future studies revealing mirror repeats in another organisms using our proposed FPCB strategy may be of interest and may also reveal various novel types of mirror repeats.

| S.No. | Gene/CDS | Sequence Reference | No of MRs | Mirror repeats |
|---|---|---|---|---|
| 1 | P53 | GenBank: AB082923.1 | 3 | 1. CCAAAGAAGAAACC<br>2. GGAACTCAAGG<br>3. ACCTGAAGTCCA |
| 2 | EIF2A | NCBI Reference Sequence NM_032025.3 | 2 | 1. ACAATTTTAACA<br>2. TTAACACAATT<br>These two mirror repeats are part of one continuous sequence ACAATTTTAACACAATT |
| 3 | STAT3 | GenBank: AJ012463.1 | 3 | 1. AGATCGGCTAGA<br>2. TCACTTTCACT<br>3. CTATCTCTATC |
| 4 | TNF-alpha | NCBI Reference Sequence X02910.1 | 3 | 1. CCTCATCTACTCC<br>2. CCAGAGGGAGACC<br>3. CCTCATCTACTCC<br>No 2 and No 3 Mirror repeats are part of one continuous sequence CCAGAGGGAGACCCCAGAGGG |
| 5 | Tgf-beta | GenBank: M60316.1 | 5 | 1. TGTGAGGGGGAGTGT<br>2. CGCCCGCGCCCGC<br>3. GCTACCACCATCG<br>4. CCGACTTCAGCC<br>5. GTCCGGGCCTG |
| 6 | Human protein kinase C theta | GenBank: L07032.1 | 2 | 1. ACATGTTTTGTACA<br>2. GACCTTTCCAG |
| 7 | CCR5) mRNA | GenBank: U54994.1 | 1 | 1. GAGAAGAAGAG |
| 8 | AKT1 | NCBI Reference Sequence NM_005163.2 | 2 | 1. AGGAGGAGGAGGA<br>2. GAGGAGGAGGAG<br>These two mirror repeats are part of one sequence AGGAGGAGGAGGAG, It contains AGG Tandem repeat and GAG Tandem repeat. |
| 9 | FOXP2 | NCBI Reference Sequence NG_007491.2 | 2 | 1. CTCTCACACTCTC<br>2. CCCAGGGACCC |
| 10 | MTOR | NCBI Reference Sequence NM_004958.3 | 4 | 1. CGCCACCACCGC<br>2. ACCTTCTTCTTCCA<br>3. ACTACAAACATCA<br>4. AGAAGAAGAAGA |
| 11 | Wnt7a | GenBank: U53476.1 | 1 | 1. GCTACGGCATCG |
| 12 | Ubiquitin | GenBank: M26880.1 | 3 | 1. CGAGGTGGAGC<br>Same sequence is repeated three times |
| 13 | BRCA1 | GenBank: L78833.1 | 5 | 1. AAGAGAAGAAAAAGAAGAGAA<br>2. AAATGAACAGACAAGTAAA<br>3. TGGATTCAAACTTAGGT<br>4. AAAGATAATAGAAA<br>5. AAACCGTGCCAAA |
| 14 | NF-kappa-B | GenBank: BC051765. | 6 | 1. CTGGGAGAGGGTC<br>2. AAAGTTATTGAAA<br>3. AAGAACAAGAA<br>4. GGAGGCGGAGG<br>5. AACGTATGCAA<br>6. GACAGTGACAG |
| 15 | COX2 | GenBank: AY462100. | 3 | 1. CGCGGTCCTGGCGC<br>2. ACAACTATCAACA<br>3. TGCAATAACGT |
| 16 | PTEN | GenBank: U93051.1 | 5 | 1. ATGATGTAGTA<br>2. AGGTTTTTGGA<br>3. AAATTTTTAAA<br>4. TTCATGTACTT<br>5. TTATAGATATT<br>First two mirror repeats are part of one continuous strand<br>ATGATGTAGTAAGGTTTTTGGA |
| 17 | HIF1A | NCBI Reference Sequence: NM_001530.3 | 7 | 1. AAAACAGTGACAAAA<br>2. TTCAGCACGACTT<br>3. GAAGACACAGAAG<br>4. GAAAAGAAAAG<br>5. CAGATTTAGAC<br>6. CCCTATATCCC<br>7. ACATTATTACA |
| 18 | PCK1 | NCBI Reference Sequence: NM_002591.3 | 2 | 1. GGTCAACAACTGG<br>2. TGGCTTTTTCGGT |
| 19 | Ceruloplasmin | GenBank: M13699.1 | 2 | 1. TCCTGGGTCCT<br>2. TAGTTTTTGAT |
| 20 | Caspase 8 | GenBank: BC068050. | 2 | 1. GTCTCACTCTG<br>2. CCTCCGCCTCC |

**Table 1 Various types of mirror repeats observed in various Human mRNA sequence using FPCB strategy**. The sequence shown in red is mirror image of sequence shown in blue.

The exact role of mirror repeats has not been understood up till date. However, as most of the genes analyzed using FPCB were found to have the presence of these repeats, a major contribution of these repeats even in the evolutionary process may be possible. Future studies regarding the detailed understanding of the functional relevance of the mirror repeats may reveal some interesting findings. We hope that our proposed FPCB strategy will be helping in these

Supplementary Data

## 1) Homo sapiens mRNA for P53, complete cds

GenBank: AB082923.1

GenBank Graphics

```
>gi|23491728:64-1245 Homo sapiens mRNA for P53, complete cds
ATGGAGGAGCCGCAGTCAGATCCTAGCGTCGAGCCCCCTCTGAGTCAGGAAACATTTTCAGACCTATGGA
AACTACTTCCTGAAAACAACGTTCTGTCCCCCTTGCCGTCCCAAGCAATGGATGATTTGATGCTGTCCCC
GGACGATATTGAACAATGGTTCACTGAAGACCCAGGTCCAGATGAAGCTCCCAGAATGCCAGAGGCTGCT
CCCCGCGTGGCCCCTGCACCAGCAGCTCCTACACCGGCGGCCCCTGCACCAGCCCCCTCCTGGCCCCTGT
CATCTTCTGTCCCTTCCCAGAAAACCTACCAGGGCAGCTACGGTTTCCGTCTGGGCTTCTTGCATTCTGG
GACAGCCAAGTCTGTGACTTGCACGTACTCCCCTGCCCTCAACAAGATGTTTTGCCAACTGGCCAAGACC
TGCCCTGTGCAGCTGTGGGTTGATTCCACACCCCCGCCCGGCACCCGCGTCCGCGCCATGGCCATCTACA
AGCAGTCACAGCACATGACGGAGGTTGTGAGGCGCTGCCCCCACCATGAGCGCTGCTCAGATAGCGATGG
TCTGGCCCCTCCTCAGCATCTTATCCGAGTGGAAGGAAATTTGCGTGTGGAGTATTTGGATGACAGAAAC
ACTTTTCGACATAGTGTGGTGGTGCCCTATGAGCCGCCTGAGGTTGGCTCTGACTGTACCACCATCCACT
ACAACTACATGTGTAACAGTTCCTGCATGGGCGGCATGAACCGGAGGCCCATCCTCACCATCATCACACT
GGAAGACTCCAGTGGTAATCTACTGGGACGGAACAGCTTTGAGGTGCATGTTTGTGCCTGTCCTGGGAGA
GACCGGCGCACAGAGGAAGAGAATCTCCGCAAGAAAGGGGAGCCTCACCACGAGCTGCCCCCAGGGAGCA
CTAAGCGAGCACTGTCCAACAACACCAGCTCCTCTCCCCAGCCAAAGAAGAAACCACTGGATGGAGAATA
TTTCACCCTTCAGATCCGTGGGCGTGAGCGCTTCGAGATGTTCCGAGAGCTGAATGAGGCCTTGGAACTC
AAGGATGCCCAGGCTGGGAAGGAGCCAGGGGGGAGCAGGGCTCACTCCAGCCACCTGAAGTCCAAAAAGG
GTCAGTCTACCTCCCGCCATAAAAAACTCATGTTCAAGACAGAAGGGCCTGACTCAGACTGA
```

Parallel complement Homo sapiens mRNA for P53

TACCTCCTCGGCGTCAGTCTAGGATCGCAGCTCGGGGGAGACTCAGTCCTTTGTAAAAGTCTGGATA
CCTTTGATGAAGGACTTTTGTTGCAAGACAGGGGGAACGGCAGGGTTCGTTACCTACTAAACTACG
ACAGGGGCCTGCTATAACTTGTTACCAAGTGACTTCTGGGTCCAGGTCTACTTCGAGGGTCTTACGG

TCTCCGACGAGGGGCGCACCGGGGACGTGGTCGTCGAGGATGTGGCCGCCGGGGACGTGGTCGG
GGGAGGACCGGGGACAGTAGAAGACAGGGAAGGGTCTTTTGGATGGTCCCGTCGATGCCAAAGGC
AGACCCGAAGAACGTAAGACCCTGTCGGTTCAGACACTGAACGTGCATGAGGGGACGGGAGTTGTT
CTACAAAACGGTTGACCGGTTCTGGACGGGACACGTCGACACCCAACTAAGGTGTGGGGGCGGGC
CGTGGGCGCAGGCGCGGTACCGGTAGATGTTCGTCAGTGTCGTGTACTGCCTCCAACACTCCGCGA
CGGGGGTGGTACTCGCGACGAGTCTATCGCTACCAGACCGGGGAGGAGTCGTAGAATAGGCTCAC
CTTCCTTTAAACGCACACCTCATAAACCTACTGTCTTTGTGAAAAGCTGTATCACACCACCACGGGAT
ACTCGGCGGACTCCAACCGAGACTGACATGGTGGTAGGTGATGTTGATGTACACATTGTCAAGGAC
GTACCCGCCGTACTTGGCCTCCGGGTAGGAGTGGTAGTAGTGTGACCTTCTGAGGTCACCATTAGAT
GACCCTGCCTTGTCGAAACTCCACGTACAAACACGGACAGGACCCTCTCTGGCCGCGTGTCTCCTTCT
CTTAGAGGCGTTCTTTCCCCTCGGAGTGGTGCTCGACGGGGGTCCCTCGTGATTCGCTCGTGACAGG
TTGTTGTGGTCGAGGAGAGGGGTCGGTTTCTTCTTTGGTGACCTACCTCTTATAAAGTGGGAAGTCT
AGGCACCCGCACTCGCGAAGCTCTACAAGGCTCTCGACTTACTCCGGAACCTTGAGTTCCTACGGGT
CCGACCCTTCCTCGGTCCCCCCTCGTCCCGAGTGAGGTCGGTGGACTTCAGGTTTTTCCCAGTCAGAT
GGAGGGCGGTATTTTTTGAGTACAAGTTCTGTCTTCCCGGACTGAGTCTGACT

## Alignment statistics for match #1

| Score | Expect | Identities | Gaps | Strand |
|---|---|---|---|---|
| 26.5 bits(28) | 0.014 | 14/14(100%) | 0/14(0%) | Plus/Minus |

```
Query   952   CCAAAGAAGAAACC   965
              ||||||||||||||
Sbjct   965   CCAAAGAAGAAACC   952
```

Range 2: 1103 to 1114GraphicsNext MatchPrevious MatchFirst Match

## Alignment statistics for match #2

| Score | Expect | Identities | Gaps | Strand |
|---|---|---|---|---|
| 22.9 bits(24) | 0.17 | 12/12(100%) | 0/12(0%) | Plus/Minus |

```
Query   1103   ACCTGAAGTCCA   1114
```

||||||||||||

Sbjct 1114 ACCTGAAGTCCA 1103

Range 3: 1044 to 1054 Graphics Next Match Previous Match First Match

Alignment statistics for match #3

| Score | Expect | Identities | Gaps | Strand |
|---|---|---|---|---|
| 21.1 bits(22) | 0.60 | 11/11(100%) | 0/11(0%) | Plus/Minus |

Query 1044 GGAACTCAAGG 1054

|||||||||||

Sbjct 1054 GGAACTCAAGG 1044

## 2) Homo sapiens eukaryotic translation initiation factor 2A, 65kDa (EIF2A), mRNA

NCBI Reference Sequence: NM_032025.3

GenBank Graphics

>gi|83656780:17-1774 Homo sapiens eukaryotic translation initiation factor 2A, 65kDa (EIF2A), mRNA

```
ATGGCGCCGTCCACGCCGCTCTTGACAGTCCGAGGATCAGAAGGACTGTACATGGTGAATGGACCACCAC
ATTTTACAGAAAGCACAGTGTTTCCAAGGGAATCTGGGAAGAATTGCAAAGTCTGTATCTTTAGTAAGGA
TGGGACCTTGTTTGCCTGGGGCAATGGAGAAAAAGTAAATATTATCAGTGTCACTAACAAGGGACTACTG
CACTCCTTCGACCTCCTGAAGGCAGTTTGCCTTGAATTCTCACCCAAAAATACTGTCCTGGCAACGTGGC
AGCCTTACACTACTTCTAAAGATGGCACAGCTGGGATACCCAACCTACAACTTTATGATGTGAAAACTGG
GACATGTTTGAAATCTTTCATCCAGAAAAAAATGCAAAATTGGTGTCCATCCTGGTCAGAAGATGAAACT
CTTTGTGCCCGCAATGTTAACAATGAAGTTCACTTCTTTGAAAACAACAATTTTAACACAATTGCAAATA
AATTGCATTTGCAAAAAATTAATGATTTTGTATTATCACCTGGACCCCAACCATACAAGGTGGCTGTCTA
TGTTCCAGGAAGTAAAGGTGCACCTTCATTTGTTAGATTATATCAGTACCCCAACTTTGCTGGACCTCAT
GCAGCTTTAGCTAATAAAAGTTTCTTTAAGGCAGATAAAGTTACAATGCTGTGGAATAAAAAAGCTACTG
CTGTGTTGGTAATAGCTAGCACAGATGTTGACAAGACAGGAGCTTCCTACTATGGAGAACAAACTCTACA
CTACATTGCAACAAATGGAGAAAGTGCTGTAGTGCAATTACCAAAAAATGGCCCCATTTATGATGTAGTT
TGGAATTCTAGTTCTACTGAGTTTTGTGCTGTATATGGTTTTATGCCTGCCAAAGCGACAATTTTCAACT
TGAAATGTGATCCTGTATTTGACTTTGGAACTGGTCCTCGTAATGCAGCCTACTATAGCCCTCATGGACA
TATATTAGTATTAGCTGGATTTGGAAATCTGAGGGGACAAATGGAAGTGTGGGATGTGAAAAACTACAAA
```

CTTATTTCTAAACCGGTGGCTTCTGATTCTACATATTTTGCTTGGTGCCCGGATGGTGAGCATATTTTAA
CAGCTACATGTGCTCCCAGGTTACGGGTTAATAATGGATACAAAATTTGGCATTATACTGGCTCTATCTT
GCACAAGTATGATGTGCCATCAAATGCAGAATTATGGCAGGTTTCTTGGCAGCCATTTTTGGATGGAATA
TTTCCAGCAAAAACAATAACTTACCAAGCAGTTCCAAGTGAAGTACCCAATGAGGAACCTAAAGTTGCAA
CAGCTTATAGACCCCCAGCTTTAAGAAATAAACCAATCACCAATTCCAAATTGCATGAAGAGGAACCACC
TCAGAATATGAAACCACAATCAGGAAACGATAAGCCATTATCAAAAACAGCTCTTAAAAATCAAAGGAAG
CATGAAGCTAAGAAAGCTGCAAAGCAGGAAGCAAGAAGTGACAAGAGTCCAGATTTGGCACCTACTCCTG
CCCCACAGAGCACACCACGAAACACTGTCTCTCAGTCAATTTCTGGGGACCCTGAGATAGACAAAAAAAT
CAAGAACCTAAAGAAGAAACTGAAAGCAATCGAACAACTGAAAGAACAAGCAGCAACTGGAAAACAGCTA
GAAAAAAATCAGTTGGAGAAAATTCAGAAAGAAACAGCCCTTCTCCAGGAGCTGGAAGATTTGGAATTGG
GTATTTAA

Parallel Complement

TACCGCGGCAGGTGCGGCGAGAACTGTCAGGCTCCTAGTCTTCCTGACATGTACCACTTACCTGGTGGTGTAA
AATGTCTTTCGTGTCACAAAGGTTCCCTTAGACCCTTCTTAACGTTTCAGACATAGAAATCATTCCTACCCTGGA
ACAAACGGACCCCGTTACCTCTTTTTCATTTATAATAGTCACAGTGATTGTTCCCTGATGACGTGAGGAAGCTG
GAGGACTTCCGTCAAACGGAACTTAAGAGTGGGTTTTTATGACAGGACCGTTGCACCGTCGGAATGTGATGA
AGATTTCTACCGTGTCGACCCTATGGGTTGGATGTTGAAATACTACACTTTTGACCCTGTACAAACTTTAGAAA
GTAGGTCTTTTTTTACGTTTTAACCACAGGTAGGACCAGTCTTCTACTTTGAGAAACACGGGCGTTACAATTGT
TACTTCAAGTGAAGAAACTTTTGTTGTTAAAATTGTGTTAACGTTTATTTAACGTAAACGTTTTTTAATTACTAA
AACATAATAGTGGACCTGGGGTTGGTATGTTCCACCGACAGATACAAGGTCCTTCATTTCCACGTGGAAGTAA
ACAATCTAATATAGTCATGGGGTTGAAACGACCTGGAGTACGTCGAAATCGATTATTTTCAAAGAAATTCCGT
CTATTTCAATGTTACGACACCTTATTTTTTCGATGACGACACAACCATTATCGATCGTGTCTACAACTGTTCTGT
CCTCGAAGGATGATACCTCTTGTTTGAGATGTGATGTAACGTTGTTTACCTCTTTCACGACATCACGTTAATGG
TTTTTTACCGGGGTAAATACTACATCAAACCTTAAGATCAAGATGACTCAAAACACGACATATACCAAAATACG
GACGGTTTCGCTGTTAAAAGTTGAACTTTACACTAGGACATAAACTGAAACCTTGACCAGGAGCATTACGTCG
GATGATATCGGGAGTACCTGTATATAATCATAATCGACCTAAACCTTTAGACTCCCCTGTTTACCTTCACACCCT
ACACTTTTTGATGTTTGAATAAAGATTTGGCCACCGAAGACTAAGATGTATAAAACGAACCACGGGCCTACCA
CTCGTATAAAATTGTCGATGTACACGAGGGTCCAATGCCCAATTATTACCTATGTTTTAAACCGTAATATGACC
GAGATAGAACGTGTTCATACTACACGGTAGTTTACGTCTTAATACCGTCCAAAGAACCGTCGGTAAAAACCTA
CCTTATAAAGGTCGTTTTTGTTATTGAATGGTTCGTCAAGGTTCACTTCATGGGTTACTCCTTGGATTTCAACGT
TGTCGAATATCTGGGGGTCGAAATTCTTTATTTGGTTAGTGGTTAAGGTTTAACGTACTTCTCCTTGGTGGAGT
CTTATACTTTGGTGTTAGTCCTTTGCTATTCGGTAATAGTTTTTGTCGAGAATTTTTAGTTTCCTTCGTACTTCGA
TTCTTTCGACGTTTCGTCCTTCGTTCTTCACTGTTCTCAGGTCTAAACCGTGGATGAGGACGGGGTGTCTCGTG
TGGTGCTTTGTGACAGAGAGTCAGTTAAAGACCCCTGGGACTCTATCTGTTTTTTTAGTTCTTGGATTTCTTCTT
TGACTTTCGTTAGCTTGTTGACTTTCTTGTTCGTCGTTGACCTTTTGTCGATCTTTTTTTAGTCAACCTCTTTTAAG
TCTTTCTTTGTCGGGAAGAGGTCCTCGACCTTCTAAACCTTAACCCATAAATT

Range 1: 467 to 478GraphicsNext MatchPrevious Match

## Alignment statistics for match #1

| Score | Expect | Identities | Gaps | Strand |
|---|---|---|---|---|
| 22.9 bits(24) | 0.38 | 12/12(100%) | 0/12(0%) | Plus/Minus |

```
Query  467  ACAATTTTAACA  478
            ||||||||||||
Sbjct  478  ACAATTTTAACA  467
```

Range 2: 454 to 467GraphicsNext MatchPrevious MatchFirst Match

Alignment statistics for match #2

| Score | Expect | Identities | Gaps | Strand |
|---|---|---|---|---|
| 21.1 bits(22) | 1.3 | 13/14(93%) | 0/14(0%) | Plus/Minus |

```
Query  643  AATAAAAGTTTCTT  656
            || |||||||||||
Sbjct  467  AACAAAAGTTTCTT  454
```

Range 3: 473 to 483GraphicsNext MatchPrevious MatchFirst Match

| Score | Expect | Identities | Gaps | Strand |
|---|---|---|---|---|
| 21.1 bits(22) | 1.3 | 11/11(100%) | 0/11(0%) | Plus/Minus |

```
Query  473  TTAACACAATT  483
            |||||||||||
Sbjct  483  TTAACACAATT  473
```

Range 4: 643 to 656GraphicsNext MatchPrevious MatchFirst Match

| Score | Expect | Identities | Gaps | Strand |
|---|---|---|---|---|

21.1 bits(22) 1.3 13/14(93%) 0/14(0%) Plus/Minus

```
Query  454  TTCTTTGAAAACAA  467
            |||||||||| ||
Sbjct  656  TTCTTTGAAAATAA  643
```